# Kinetics of photo-dissolution within Ag/As$_2$S$_3$ heterostructure

Pritam Khan[1,2,3], Yinsheng Xu[2], William Leon[4], K. V. Adarsh[3], Dmitri Vezenov[4], Ivan Biaggio[5] and Himanshu Jain[2*]

[1]Physics, Kyushu University, 744 Motooka, Fukuoka, 819–0395, Japan

[2]Materials Science and Engineering, Lehigh University, Bethlehem, 18015, PA, USA

[3]Physics, IISER Bhopal, Bhopal Bypass Road, Bhauri, Bhopal, 462066, Madhya Pradesh, India

[4]Chemistry, Lehigh University, Bethlehem, 18015, PA, USA 6Physics, Lehigh University, Bethlehem, 18015, PA, USA

[5]Physics, Lehigh University, Bethlehem, 18015, PA, USA 6Physics, Lehigh University, Bethlehem, 18015, PA, USA

*Chalcogenide glass – silver heterostructures are candidates for photoresist and diffractive optical applications. To optimize their processing, we report the kinetics of Ag photo-dissolution in As$_2$S$_3$ matrix using in-situ optical transmission/reflection measurements and real time atomic force microscopy (AFM) imaging under optical illumination. The results indicate that photo-dissolution occurs in three stages with the extent and kinetics of each stage depending strongly on Ag film thickness. By contrast, the photo-dissolution is found to be independent of As$_2$S$_3$ matrix thickness. The extent of three stages also depends strongly on the laser dose and can be reduced to two stages at higher laser fluence. A comparative study of two oppositely stacked sample configurations: As$_2$S$_3$/Ag/glass and Ag/As$_2$S$_3$/glass show that the heterostructures respond differently to light illumination. For the former, Ag dissolves completely into As$_2$S$_3$ matrix at a faster rate than for the latter case. The origin of this difference is established by energy dispersive X-ray spectroscopy and AFM measurements.*

**Keywords:** Thin films; photo-dissolution, chalcogenides, pump-probe, spectroscopy

# 1. Introduction

Heterostructures built from multiple components often exhibit physical-chemical properties that are not feasible with either of the individual components [1, 2]. Silver-chalcogenide glasses (ChG) bilayers, such as $As_2S_3$/Ag, are of particular interest for the formation of multilayer heterostructures because of their potential applications in optics and optoelectronics [3, 4] An immediate application could be the fabrication of Programmable Metallization Cell (PMC) memory devices. Their formation requires the dissolution of metallic Ag into ChG matrix. Recently, Kozicki *et al*. [5] have shown that a solid electrolyte can be formed when 33 atomic % Ag is dissolved in $Ge_{30}Se_{70}$ [6]. Furthermore, the high transparency of ChGs in the IR region in combination with a variety of photo-induced effects can be used for making optical devices [6, 7]. For example, in $As_2S_3$/Ag bilayer, photo-induced dissolution of metal Ag has attracted much interest for fabricating IR diffractive elements because it produces large changes in the physical and optical properties of glass, particularly its refractive index and etch resistance [8]. The dependence of dissolution rate on optical illumination renders these materials suitable for photoresists for lithography [9]. Upon irradiation with bandgap light, Ag and $As_2S_3$ layers start interacting with each other, changing the chemical composition as well as the structure of the irradiated region compared to the unirradiated region [9]. Such photo-induced effects from the photo generation of electron-hole pairs alter the transmission, reflection, and micro-hardness of the heterostructure. Due to the difference in electron and hole mobility, holes diffuse away from the illuminated region faster than the electrons [10, 11]. As a result, the negatively charged illuminated region attracts metal ions from the surface, complementing ion diffusion with a strong drift component [12]. However, the details of the underlying processes remain unclear. An understanding of the kinetics of Ag photo-dissolution in $As_2S_3$ can help us in optimizing the performance of these materials for photoresist and other applications.

Since the discovery of photo-dissolution by Kostyushin *et al*. [13], the time dependence of physical properties such as optical transmittance/reflectance [14] or electrical resistivity [15], and the concentration profile of Ag, have been determined [16]. Although a detailed explanation of the dissolution phenomenon is still being debated, it is generally believed that Ag photo-dissolution consists of three stages: (a) an induction period; (b) the effective photo-dissolution, and (c) a final stage that completes the photo-dissolution [16, 17]. For practical applications, it is important to delineate these stages and also establish how they may be affected by the stacking sequence of these layers, especially in the context of multilayer structures. Accordingly, we have studied the kinetics of photo-dissolution in two oppositely stacked heterostructures of Ag and $As_2S_3$ films: $As_2S_3$/Ag/glass and Ag/$As_2S_3$/glass (from top to bottom). We have chosen $As_2S_3$ as the ChG matrix layer because sulfides have a higher ionic activity, which is favorable for investigating photo-dissolution mechanisms as well as practical applications. To obtain insights of the photo-dissolution process, we have varied the thickness of both Ag and $As_2S_3$ layers in either configuration. We have also performed real time atomic force microscopy (AFM) imaging under simultaneous optical illumination to obtain direct evidence of the photo-dissolution process at the nanoscale.

## 2. Experimental

Bilayer films of $As_2S_3$ ChG matrix and metallic Ag were deposited on a microscopy glass slide by conventional thermal evaporation in a vacuum of $1\times10^{-6}$ mbar using commercially available bulk materials (5N purity). A low deposition rate of 5–8 Å/s was maintained and continuously monitored using a quartz crystal. It is well known that such a low deposition rate produces thin films of composition that is very close to the starting bulk materials [18-21]. We prepared bilayer films in two different stacking configurations: $As_2S_3$/Ag/glass and Ag/$As_2S_3$/glass as shown in Fig.1, where the absorption coefficient of $As_2S_3$ is $4.2 \times 10^3$ cm$^{-1}$. We prepared three sets of samples: (1) with varying thickness of Ag layer (5, 10, 20 and 40 nm) for a fixed thickness of a-$As_2S_3$ matrix (Set 1), (2) with varying thickness of $As_2S_3$ matrix (430 and 860 nm) with fixed Ag layer thickness (Set 2), and (3) with opposite stacking sequence, where Ag is deposited at the bottom or the top of a-$As_2S_3$ (Set 3). The three sets of samples are listed in Table 1. The optical band gap ($E_g$) of all samples was calculated using classical Tauc plot [22]. We found that $E_g$ varies between 2.38 eV (521 nm) to 2.40 eV (516 nm) for the samples used in the present study. Therefore, 488 nm (2.54 eV) laser provided near bandgap irradiation to the samples. We have used computer program "PARAV" for calculating the refractive index of all samples from experimentally measured transmission spectra of thin films [23]. The refractive index values of samples are shown in Table 1.

We have employed *in-situ* pump-probe optical transmission/reflection measurements. The bilayer is illuminated with a continuous wave 488 nm Ar$^+$ laser as pump beam with 4 mm spot size of intensity 7.5

| Ag (nm) | As$_2$S$_3$ (nm) | Ag (position) | Set | Bandgap (eV) | Refractive index |
|---|---|---|---|---|---|
| 5 | | | Set 1 | 2.40 ± 0.002 | 2.38 |
| 10 | 430 | Bottom | | 2.40 ± 0.003 | 2.42 |
| 20 | | | | 2.39 ± 0.002 | 2.46 |
| 40 | | | | 2.38 ± 0.001 | 2.56 |
| 20 | 430 | Bottom | Set 2 | 2.39 ± 0.002 | 2.46 |
| | 860 | | | 2.38 ± 0.003 | 2.68 |
| 20 | 430 | Bottom | Set 3 | 2.39 ± 0.002 | 2.46 |
| | | Top | | 2.38 ± 0.001 | 2.66 |
| 20 | 860 | Bottom | Set 3 | 2.38 ± 0.003 | 2.68 |
| | | Top | | 2.38 ± 0.002 | 2.78 |

Table 1. Different Sets of samples to study of Ag photo dissolution within As$_2$S$_3$ matrix. Set 1: Ag thickness is varied and is deposited below As$_2$S$_3$ of fixed thickness 430 nm. Set 2: 20 nm Ag is deposited below As$_2$S$_3$ whose thickness is kept at 430 and 860 nm. Set 3: Combination of Set 1 and 2, where 20 nm Ag is deposited top and bottom of 430 and 860 nm As$_2$S$_3$.

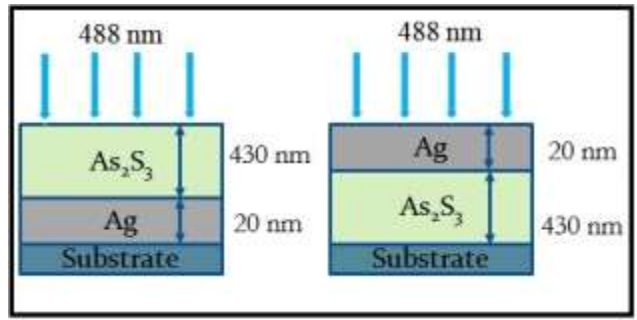

Fig. 1. Schematics of the heterostructure in two different geometric configurations: As$_2$S$_3$/Ag/glass and Ag/As$_2$S$_3$/glass. The blue arrows show the direction of laser illumination.

mW/cm$^2$, and a very weak white light beam (400 to 900 nm) as probe with a very low intensity and a smaller diameter of 2 mm. The two beams are directed in such a way that they cross each other on the sample surface. The transmission/reflection through/from the sample is then measured by detecting the transmitted or reflected white light probe with a fiber optic Ocean Optics spectrometer, which has the ability to collect the entire spectrum in 2 ms. A schematic of the experimental set up is shown in Fig. S1 of Supplemental Information (SI). Note that for all measurements, the pump beam illuminated the top surface of the film. To assess the influence of light intensity on dissolution, the intensity of the pump beam is varied from 7.5 to 0.5 mW/cm$^2$. It is noteworthy that all the pump-probe measurements were performed under ambient condition.

AFM imaging was performed using an Asylum Research MFP-3D Bio atomic force microscope paired with an Olympus IX-71 inverted optical microscope. Illumination was provided by an X-Cite Series 120 broadband light source filtered with an Olympus Z473/10x excitation filter (peak intensity 477 nm, FWHM 10 nm) and focused onto the sample plane with a 10x objective. The sample in this setup was illuminated from below and AFM imaging done from above. Power density in the sample plane during imaging was 1.4 mW/cm$^2$. Topographic images in air were obtained using a Budget Sensors Tap300Al cantilever in AC mode. Image scan size was set to 1 µm x 1 µm (at 256 x 256 pixels) and taken at a scan rate of 2.1 Hz (requiring 121 s per image frame). The EDAX measurements were performed using Oxford INCA system with an X-Max detector.

## 3. Results and Discussions

*3.1 Photo-induced optical changes*

To extract the contributions of individual layers, we measured the optical transmittance (T) of a pure Ag film and a pure a-As$_2$S$_3$ on a glass slide, and compared them with that of the bilayer heterostructure, as shown in Fig.2(a). Note that in the as-prepared state before the pump beam is switched on (t = 0) the transmission spectra of the bilayer are a composite of contributions from pure Ag and a-$As_2S_3$, leading to a typical modulation in transmission that is caused by the interference of multiple reflections at the interfaces. These interference fringes contain a wealth of information about the structure of the deposited films and how they evolve in time. Notably, the bilayer spectrum exhibits similar interference fringes like a-$As_2S_3$ but modulated by the envelope (maxima/minima) of the transmission spectrum of pure Ag. With time the modulation part gradually diminishes, so that at the final stage of light illumination, the spectrum looks qualitatively similar to that of the pure $As_2S_3$ spectrum, except that it has shifted significantly to longer wavelengths. This evolution of the transmission spectra indicates that eventually Ag completely dissolves into the matrix and results in an optically homogeneous a-(Ag+$As_2S_3$) film.

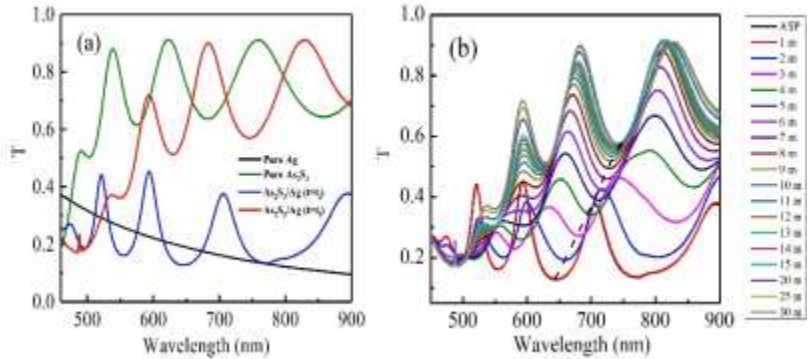

Fig. 2. (a) Transmission spectra of metallic Ag, pure $As_2S_3$ and bilayer composite $As_2S_3$/Ag at the initial and final states of laser illumination. In the as-prepared state (t=0), the transmission spectrum of the bilayer composite is a combination of metallic Ag and pure $As_2S_3$. In the final state, the spectra look similar to that of pure $As_2S_3$ indicating that entire Ag has dissolved into $As_2S_3$ matrix. (b) Time evolution of transmission spectra of the sample $As_2S_3$/Ag ($As_2S_3$=430 nm, Ag=20 nm) at different stages of illumination. Note that the spectra red shift (black dotted line) and the sample becomes more transparent as compared to the initial state.

To identify the stages of photo-dissolution, the transmission spectrum of the bilayer was recorded continuously in the presence of the pump beam illumination. A typical transmission curve is shown in Fig.2(b) as a function of exposure time for one of the samples. The small spike at 488 nm is scattered light from the pump beam. The transmission spectrum is characterized by strong absorption edge followed by weekly absorbing interference fringes at longer wavelengths. To consistently quantify the results and correlate with the kinetics of photo-dissolution, we have focused on the second minimum of the transmission spectra after the absorption edge. The time-evolution of this minimum delivers two important observations: (1) The transmission minimum shifts significantly towards longer wavelength, as indicated by the black dotted line connecting all minima at different times, and (2) the magnitude of minimum gradually increases, indicating that the film becomes increasingly more transparent with laser illumination.

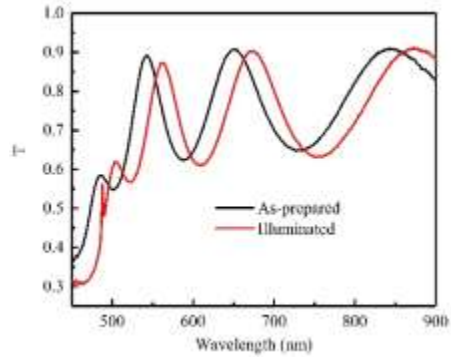

Fig. 3. Transmission spectra of pure $As_2S_3$ of thickness 430 nm without any doped Ag in as-prepared and illuminated state.

To establish the effect of Ag photo-dissolution unambiguously, we first recorded the pump-probe data for pure $As_2S_3$ without any Ag layer. As seen in Fig.3, the transmission spectra exhibit photodarkening, shifting to longer wavelengths, with little to no change in the magnitude of the transmission at longer wavelengths. Next, we performed pump-probe measurements on the samples with stacking sequence: $As_2S_3$/Ag/glass (Set 1). The resultant transmission spectra of all samples are shown in supplementary Fig. S2 for a pump intensity of 7.5 mW/cm². The time evolution of the second transmission minimum extracted from these data is obtained and plotted in Fig.4(a) for samples of varying Ag thickness (5, 10, 20 and 40 nm). As the Ag film thickness decreases, the transmission curves shift upwards, as expected because of the lower reflectivity of the thinner Ag layers before photo-dissolution in the chalcogenide layer. Further, with decreasing Ag thickness the amount of material to dissolve decreases, and the photo-dissolution process becomes faster: the transmission curves in Fig.4(a) become steeper and gradually move to the left, as indicated by the red dotted horizontal line.

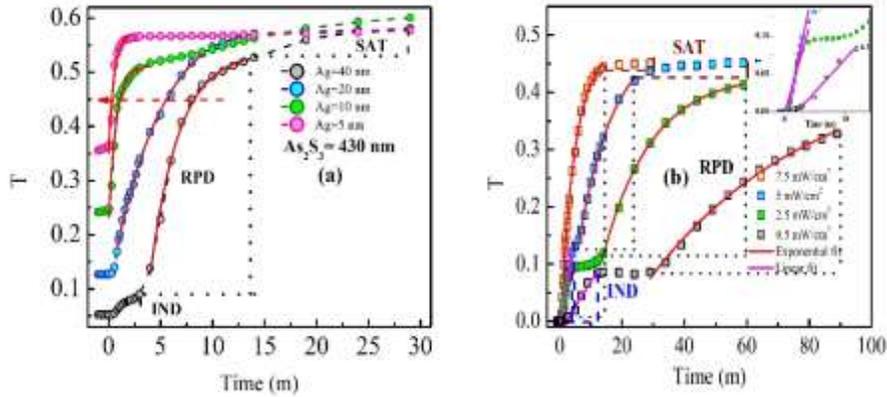

Fig. 4. (a) Time evolution of the height of the second transmission minimum for Set 1 samples with Ag film thickness of 5, 10, 20 and 40 nm, and fixed thickness of $As_2S_3$ matrix at 430 nm. The dots define the different stages of photo-dissolution for the sample Ag=40 nm. The solid red line represents a single exponential build-up $(1 + \exp(-t/\tau))$ that fits the rapid photo-diffusion (RPD) part of photo-dissolution. (b) Temporal evolution of transmission minimum height for four different pump beam intensity for the sample Ag=20 nm. The blue and brown dashed lines defined the IND and SAT stages respectively. The region indicated by black dots represents RPD stage. The solid red line represents theoretical fitting of the RPD and IND stages, respectively, to single exponential $(1 + \exp(-t/\tau))$ function. The inset in (b) shows the linear fitting as solid pink line corresponding to the IND stage. Sample with x=5 is not shown as it does not have any IND stage.

For all samples the final transmission saturates at about the same value, which indicates that irrespective of thickness, ultimately Ag dissolves completely into $As_2S_3$ matrix without significantly increasing its absorption. However, the initial dissolution rate increases with the intensity of light at the Ag-As$_2$S$_3$ interface, which itself is enhanced with decreasing Ag film thickness.

Before discussing our results further, we note that to date, many experiments have been performed and several models have been proposed to describe the mechanisms of Ag photo dissolution. Kolobov and Elliott [24] published a review and concluded that the photo-dissolution process comprises of three stages: (a) Induction period, (b) Effective dissolution and (c) Final stage of photo-dissolution. Elliott [25] proposed a unified mechanism for metal photo-dissolution in amorphous chalcogenide materials, based on the combined ionic and electronic charge-carrier transport. Later, Jain *et al*. [26] investigated the very early stage of photo-diffusion at the Ag/As$_2$S$_3$ interface and proposed a two-stage mechanism for the initiation of metallic silver to diffuse as silver ions. Subsequently, Popescu *et al*. [17, 27] performed several studies to understand the different stages of photo-dissolution.

These different stages are most visible in the spectrum for the sample with 40 nm Ag in Fig.4(a), which shows qualitative differences compared to the remaining three samples with smaller thicknesses. In fact, our experimental curve in Fig.4(a) is most similar to what was observed by Sava *et al*. [28]. Here, photo-dissolution kinetics is not linear, but consists of different stages, each of which is identified by a characteristic slope in the rate of transmission change. In this description, photo-diffusion within the 40 nm Ag film sample consists of three distinct stages vs. two stages for the thinner samples (see Fig.4(a)). In the case of the thinner samples, T increases rapidly as soon as the pump beam is turned on at t = 0, corresponding to a rapid photo-diffusion (RPD) stage. Then, the rate of transmission gradually decreases, forming a plateau at the saturation (SAT) stage that is due to the exhaustion of Ag when it is fully distributed within the ChG matrix. On the other hand, for the thicker sample, the increase in T is relatively slow (i.e. the Ag photo-dissolution rate is slow) during the approximate time interval of 0 - 3 minutes in Fig.4(a), appearing to reach a brief plateau. This stage of photo-dissolution is called the induction (IND) stage, which was firstly observed by Goldschmidt and Rudman [15]. It is striking that we did not observe any IND stage for thinner Ag samples. At this point, it is important to note that the non-observation of IND stage does not necessarily mean that it is absent. In fact, a reasonable argument could be that RPD stage appears so fast that it masks IND stage.

The IND stage is characterized by the time Ag atoms take to become chemically reactive. It has been proposed that structural interface defects, which change the local electric field of unsatisfied bonds, are responsible for local activation and reaction sites [28]. Then, in the next stage the transmission increases rapidly, much like for the samples with thinner Ag layer – this stage is characterized by a high Ag dissolution rate and is the RPD stage that is the main part of Ag photo-dissolution. The duration of this photo-dissolution stage depends strongly on the amount of Ag available and also on the laser dose i.e. the product of light intensity and illumination time which we will discuss below. To describe it quantitatively, we have fitted the RPD part of the transmission curve (red solid line in Fig.4(a)) for all samples with a single exponential growth function proportional to $(1 + \exp(-t/\tau))$. The resulting exponential time constants are shown in Table 2.

| Ag thickness (nm) | Time constant (min) | Squared correlation coefficient |
|---|---|---|
| 5 | $0.5 \pm 0.09$ | 0.965 |
| 10 | $0.7 \pm 0.06$ | 0.977 |
| 20 | $2.8 \pm 0.23$ | 0.993 |
| 40 | $4.8 \pm 0.22$ | 0.998 |

Table 2. Time constants for RPD stage of photo-dissolution. Clearly, RPD stage becomes faster when the Ag thickness decreases.

Clearly, the RPD stage saturates faster when Ag thickness decreases, which is expected as the amount of Ag is less and it dissolves faster into the $As_2S_3$ matrix.

The photo-dissolution of Ag in $As_2S_3$ can be considered as a photo-induced solid-state reaction between the metal layer and the chalcogenide [29]. It is characterized by a highly complex electron and ion transfer and transport processes. According to Kluge [29], the RPD stage is controlled by a chemical reaction at the interface between Ag-doped and undoped $As_2S_3$ layers, which has been studied experimentally in detail by many authors including Jain *et al.* [26] and Popescu *et al.* [27]. In short, upon light illumination, through the intermediation of homopolar S–S bonds an electric double layer consisting of positively charged $Ag^+$ ions and a negatively charged As–S layer is developed at the interface of the illuminated region. Then laser illumination helps to detach and transfer the Ag into the $As_2S_3$ matrix as $Ag^+$ ions. Continuous illumination causes a charge gradient, which then produces a preferential drift direction for the roaming $Ag^+$ ions, pushing them away from the interface, toward the inside of the $As_2S_3$ film [12, 28] Thus the last part, i.e. the SAT stage is associated with the completion of the movement of $Ag^+$ ion into ChG matrix. According to Sava *et al.* [28] one possibility is that inside $As_2S_3$ matrix, the $Ag^+$ ion could be sucked into structural voids with an appropriate radius. Possible examples of such regions include Ag that makes three bonds with S in the $As_2S_3$ matrix, as suggested by Steel *et al.* [30], Penfold and Salmon [31] and Kovalskiy *et al.* [32] from EXAFS measurements. As a result, as these sites become filled photo-dissolution slows down and finally saturates.

To understand the origin of the IND stage observed for the thick sample of Fig. 4(a), we have studied the dissolution kinetics of a sample with relatively thin Ag (20 nm) film, and a reduced pump intensity from 7.5 to 0.5 $mW/cm^2$. The results, which are shown in Fig.4(b), indicate that the kinetics of photo-dissolution becomes slower and saturation takes a longer time at lower pump intensities. Most importantly, the IND stage reappears at all intensities lower than 7.5 $mW/cm^2$ with strong intensity dependent kinetics. The kinetics of the optical transmission in the IND stage can be approximated with a linear time-dependence, as shown in the inset of Fig.4(b). The linear fits become steeper with increasing pump intensity as indicated by the slopes of the linear sections in Table 3.

| Intensity (mW/cm$^2$) | Slope (min$^{-1}$) | Pearson correlation coefficient |
|---|---|---|
| 0.5 | 0.009 ± 5.19 x 10$^{-4}$ | 0.991 |
| 2.5 | 0.02 ± 1.9 x 10$^{-3}$ | 0.987 |
| 5 | 0.02 ± 1.5 x 10$^{-3}$ | 0.979 |

Table 3. Slopes of the IND stage obtained for different intensities of the laser. Note that the slope becomes steeper, i.e. IND stage becomes faster at higher intensities of the laser beam.

Table 1 shows that the slope of the curve becomes gradually steeper, i.e. IND stage progresses faster with increasing intensity. After the IND stage, the RPD stage commences, which also depends strongly on excitation intensity. However, the RPD part of the transmission kinetics curve follows a single exponent function $(1 + \exp(-t/\tau))$. Its time constant for various intensities are given in Table 4.

| Intensity (mW/cm$^2$) | Time constant (min) | Squared correlation coefficient |
|---|---|---|
| 0.5 | 52 ± 4.99 | 0.995 |
| 2.5 | 15 ± 1.85 | 0.993 |
| 5 | 9 ± 0.49 | 0.996 |
| 7.5 | 4 ± 0.13 | 0.998 |

Table 4. Exponential time constants obtained for RPD stage of photo-dissolution at different intensities of the laser.

From Table 4, we find that the kinetics of RPD stage becomes ten times faster when laser intensity is increased from 0.5 to 7.5 mW/cm$^2$. At 2.5 and 0.5 mW/cm$^2$, we could not observe the SAT stage within our experimental time frame as the kinetics were significantly slowed down. Our studies of the kinetics of the optical transmission reveal that the duration of both RPD and IND stages is roughly inversely proportional to the light intensity. In addition, the relative importance of the different stages of dissolution depends on the pump intensity. For example, at the maximum intensity of 7.5 mW/cm$^2$ we observed only two stages of dissolution, with the IND stage completely absent. For 5 mW/cm$^2$, we could observe all three stages of dissolution. At the lower intensities of 2.5 and 0.5 mW/cm$^2$, we also observed the two IND and RPD stages.

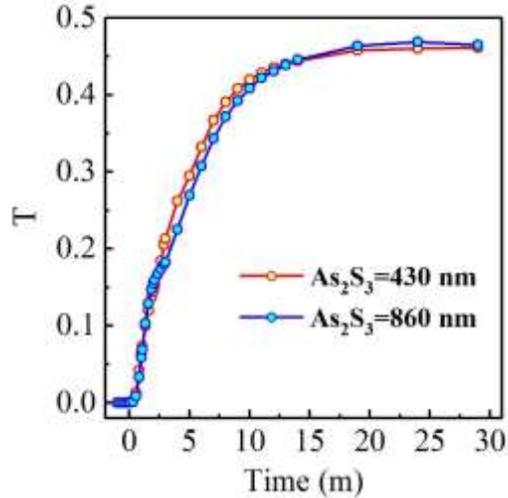

Fig. 5. Time evolution of transmission minimum for two values of As$_2$S$_3$ thickness (430 and 860 nm) but fixed Ag film thickness at 20 nm. The solid lines are drawn as guide to eyes.

Having observed a complex impact of Ag thickness on the photo-dissolution kinetics, it is natural to ask if the thickness of the As$_2$S$_3$ matrix also affects the kinetics. We therefore performed in-situ pump-probe transmission measurements on the samples of Set 2, where the thickness of Ag layer was fixed at 20 nm, while the As$_2$S$_3$ layer thickness was doubled from 430 nm to 860 nm. The resulting transmission spectra are shown in supplementary Fig. S3 and the corresponding kinetics of the transmission minimum are plotted in Fig.5 for the two values of the As$_2$S$_3$ layer thickness. The results are practically indistinguishable from each other, suggesting that the thickness of As$_2$S$_3$ layer plays little to no role in photo-dissolution process. Such results indicate that diffusion speed of Ag ion in the ChG matrix is not a limiting factor, i.e. it is much faster than the photo-dissolution process.

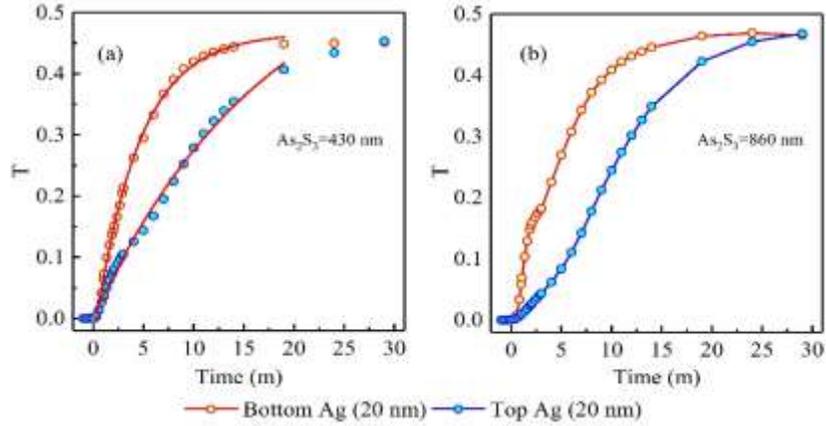

Fig. 6. Dependence of Ag photo-dissolution as observed by change in the transmission through $As_2S_3$/Ag/glass and Ag/ $As_2S_3$/glass stacking sequences. Thickness of Ag film is 20 nm, whereas the thickness of $As_2S_3$ film is (a) 430 nm and (b) 860 nm. The red lines in (a) represent the exponential fit following $(1 + \exp(-t/\tau))$, whereas the solid lines in (b) are drawn as guide to eyes.

Illuminating the stacked layers from the ChG side or from the Ag side will change the spatial distribution of the pump light intensity in the sample, which should have an influence on the photo-dissolution process. To assess this prediction, we have compared the response of two oppositely stacked Ag and $As_2S_3$ films in the heterostructures: $As_2S_3$/Ag/glass and Ag/$As_2S_3$/glass (Set 3 in Table 1). Because of the deposition process, there is also a difference in the morphology of the Ag/ChG interface between the two samples. The Ag/ChG interface is flat and featureless when the Ag layer is at the bottom, but the Ag layer with a nominal thickness of 20 nm that was deposited on top of the ChG aggregated into 100 nm islands (See below). The as-obtained transmission spectra are shown in supplementary Fig. S4. The time evolution of transmission spectra for the two configurations are calculated as before. The results are shown in Fig.6(a) and (b) for the two types of heterostructures and two thicknesses of the $As_2S_3$ layer. The main outcomes of the data analysis are:

(1) The kinetics of the RPD state of photo-dissolution is faster when Ag is deposited below the $As_2S_3$ layer compared to when it is deposited on the top. This is due to the fact that less of the incident light is absorbed in the Ag layer, with the pump light directly illuminating the Ag/ChG interface. A single exponential fit (1 + exp (-t/$\tau$)) to the data (solid red line) in Fig.5a determined exponential time constants of 4.4 ± 0.1 and 17.6 ± 2.1 minutes for illumination through the ChG matrix or through the Ag layer, respectively.

(2) The onset of SAT stage takes place faster in bottom-deposited Ag samples compared to when it is deposited on the top, because of the difference in pump light intensity at the interface. As a result, in the latter case, we could not witness saturation of the kinetics curve and complete dissolution of Ag within our experimental time window of 30 minutes, which was otherwise observed for the former. The faster kinetics obtained for Ag-bottom geometry can be explained by the fact that for the Ag-top geometry a significant fraction of the light is reflected form the Ag surface and only a fraction reaches the Ag/$As_2S_3$ interface to trigger the dissolution process. On the other hand, for the Ag-bottom configuration, the $As_2S_3$/Ag interface receives a much greater fraction of light since the reflectivity of $As_2S_3$ is lower than that of the Ag film, which in turn accelerates the dissolution process and consequently accelerates the changes in transmission. To substantiate this explanation, we have performed in-situ pump-probe reflection measurements on the samples with Ag deposited on top of $As_2S_3$ matrix. The resulting evolution of reflection spectra is shown in Fig.7(a) and (b). It is clear that for both samples reflectivity is maximum in the initial state before exposure to pump laser. Upon laser illumination, reflected intensity decreases as Ag gradually dissolves in $As_2S_3$ matrix. In addition, the reflection spectra also red shift with time, similar to what was observed from transmission measurements.

## 3.2 Photo-induced structural changes

To determine the extent of Ag photo-dissolution, we have performed energy dispersive X-ray analysis (EDAX) measurements on the same sample before and after laser illumination as shown in Fig.7(c) and (d). The EDAX spectrum for the as-prepared sample shows the presence of As, S and Ag. To analyze the concentrations semi-quantitatively, we have calculated the peak intensity ratio of constituent components in both as-prepared and illuminated state. For example, both the Ag/As ratio and the Ag/S ratio decreased between the as-prepared state and the post-illuminated state (a decrease of 0.22 to 0.15 for Ag/As and from 0.53 to 0.37 for Ag/S). Such observations indicate that Ag density decreases with illumination as it dissolves into the $As_2S_3$ matrix. Importantly, As/S ratio remains unchanged at 2.45 in both as-prepared and post-illuminated state. This observation indicates that during the entire period of illumination there is no significant migration of As or S near the surface of the sample. However, different bonds or structural units may still break and form because of illumination.

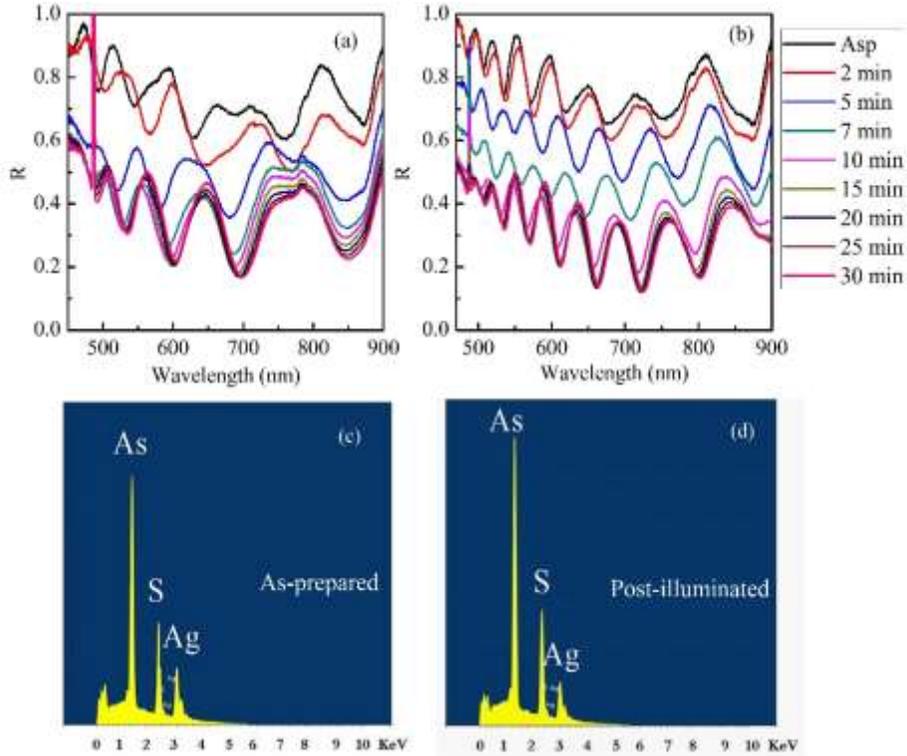

Fig. 7. Time resolved reflection of spectra of heterostructure with 20 nm Ag deposited on top of (a) 430 and (b) 860 nm $As_2S_3$ matrix. Reflectivity of the samples is the largest in the as-prepared state, which decreases gradually with time upon laser illumination. EDAX spectrum of $Ag/As_2S_3$ (20/430 nm) bilayer in (c) as-prepared, and (d) post-illuminated states.

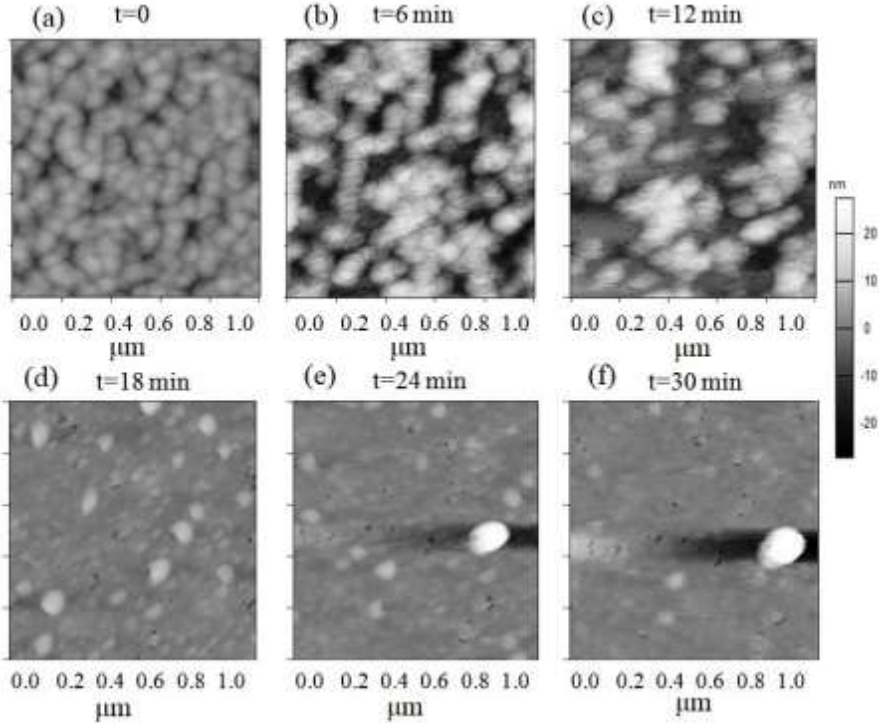

Fig. 8. Topographical AFM image of the heterostructure which shows the dissolution of Ag into $As_2S_3$ matrix.

To get direct evidence of kinetics of Ag photo-dissolution in $As_2S_3$ matrix, we performed in-situ real time AFM imaging of the Ag/$As_2S_3$/glass heterostructure. Fig.8(a) shows a topographical AFM image of the sample in the as-prepared state, which reveals a uniform nanoparticulate nature of Ag film on the $As_2S_3$ matrix, similar to that reported in the literature [33]. The average size of Ag nanoparticles is about 100 nm, which are uniformly distributed on the surface of $As_2S_3$ matrix. To investigate the kinetics of Ag photo-dissolution, a series of AFM images was recorded continuously throughout the period of illumination. Each image represents the same dimension of the probed area (1μm x 1μm) with the image capturing time of 2 minute/frame. The progression of Ag dissolution can be seen in figs. 8(b) to (f) from time resolved changes in the surface morphology of the heterostructure. The black-white vertical contrast is kept constant at ±20 nm. From these images we find that upon light illumination Ag starts dissolving in the $As_2S_3$ matrix. It is evident from Fig.8 that with the progression of time, Ag moves sideways and dissolves into the $As_2S_3$ matrix. The dissolution rate of the Ag layer is very fast for the initial period of illumination. It can be seen from Fig.7 that the density of Ag nanoparticles on the surface in Fig.7d is much less than that of Fig.8(a). It indicates that by 18 minutes most of Ag has dissolved into $As_2S_3$ matrix. At later times, i.e. t=24 and 30 minutes the AFM images doesn't shown any appreciable change in the Ag density except that an artifact appeared in the right-hand side of the images. At this point, our result is somewhat different from the observation of Palumbo *et al* [33], where they observed the re-emergence of Ag islands in the post illumination state. In the present case, we did not observe the reappearance of new nanostructures during the period of illumination at longer times, but we could observe some leftover Ag on the surface of the sample. It indicates that within our experimental time window of t=30 min, Ag does not dissolve completely into $As_2S_3$ matrix. This observation supports the kinetic data that transmission through the heterostructure does not saturate within t=30 minutes (Fig.6) when Ag is deposited on top of $As_2S_3$ layer.

## 4. Conclusions

The kinetics of Ag photo-dissolution into an $As_2S_3$ matrix has been studied using in situ pump-probe spectroscopy and real time AFM imaging. Kinetic analysis reveals that dissolution occurs in three stages: induction (IND), rapid photo-diffusion (RPD) and saturation (SAT). The transmission in the IND increases rapidly to quickly saturate at an intermediate level, whereas in the RPD stage it is characterized by a single exponential build-up until the SAT stage is reached. The duration of each stage depends strongly on laser intensity, as well as on the thickness of the Ag layer. For example, the IND stage for thin Ag samples becomes too short to be resolved experimentally.

The dissolution process in the heterostructure of two opposite geometric configurations – $As_2S_3$/Ag/glass and Ag/$As_2S_3$/glass produce different results, with the photo dissolution process faster for the $As_2S_3$/Ag/glass sample because of the higher laser intensity at the Ag-$As_2S_3$ interface. The different morphology of the Ag-$As_2S_3$ interface also played a role. Real time AFM imaging during optical illumination provides direct evidence of photo-dissolution.

**Acknowledgement**

H. J., K. V. A. and P. K. thank the National Science Foundation through the International Materials Institute for New Functionality in Glass (DMR-0844014) for supporting our international collaboration between Lehigh University, USA, and Indian Institute Science Education and Research, Bhopal, India. P. K. and K. V. A. thank Department of Science and Technology (Project no: SR/S2/LOP-003/2010) and Council of Scientific and Industrial Research, India, (grant No. 03(1250)/12/EMR-II) for supporting the work at IISER.

# References

(1) P.-J. Wu, J.-W. Yu, H. J. Chao and J.-Y. Chang, Chem. Mater. **26**, 3485 (2011).

(2) P.Němec, J.Charrier, M.Cathelinaud, M.Allix, J.-L.Adam, S.Zhang, and V.Nazabal, Thin Solid Films **539**, 226 (2013).

(3) M. Bapna, R. Sharma, A. R. Barik, P. Khan, R. R. Kumar, and K. V. Adarsh, Appl. Phys. Lett. **102**, 213110 (2013).

(4) W. Leung, N. W. Cheung, and A. R. Neureuther, Appl. Phys. Lett. **46**, 481 (1985).

(5) M. Mitkova, and M. N. Kozicki, J. Non-Cryst. Solids **299**, 1023 (2002).

(6) T. Wagner, and P. J. S. Ewen, *J. Non-Cryst. Solids* **266**, 979-984 (2000).

(7) E. Baudet, A. Gutierrez-Arroyo, P. Němec, L. Bodiou, J. Lemaitre, O. De Sagazan, H. Lhermitte, E. Rinnert, K. Michel, B. Bureau, J. Charrier, and V. Nazabal, Opt. Mater. Express **6**, 2616 (2016).

(8) T. Wagner, E. Marquez, J. F. -Pena, J. M. G. - Leal, P. J. S. Ewen and S. O. Kasap, Philos. Mag. B **79**, 223 (1999).

(9) H. Jain and M. Vlcek, J. Non-Cryst. Solids **354**, 1401 (2008).

(10) S. Binu, P. Khan, A. R. Barik, R. Sharma, R. Golovchak, H. Jain and K. V. Adarsh, Mater. Res. Express **1**, 045025 (2014).

(11) T. Kawaguchi and S. Maruno, J. Appl. Phys. **77**, 628 (1995).

(12) I. Kaban, P. J´ov´ari, T. Wagner, M. Frumar, S. Stehlik, M. Bartos, W. Hoyer, B. Beuneu and M. A. Webb, J. Phys.: Condens. Matter **21**, 395801 (2009).

(13) M. T. Kostyshin, E. V. Mikhailovskaya, P. F. Romanenko and F. T. Tela, Soviet Phys. Solid State **8**, 571 (1966).

(14) A.P. Firth, P.J.S. Ewen and A.E. Owen. J. Non-Cryst. Solids **77-78**, 1153 (1985).

(15) D. Goldschmidt and P.S. Rudman. J. Non-Cryst. Solids **22**, 229 (1976).

(16) Y. Yamamoto, T. Itoh, Y. Hirose and H. Hirose. J. Appl. Phys. **47**, 3603 (1976).

(17) M. Popescu, F. Sava, A. Lorincizi, A. Velea, M. Leonovici, S. Zamfira, J. Optoelectron. Adv. M. **11**, 1586 (2009).

(18) P. Khan, R. Sharma, U. Deshpande, and K. V. Adarsh, Opt. Lett. **40**, 1559 (2015).

(19) P. Khan, P. Acharja, A. Joshy, A. Bhattacharya, D. Kumar, and K.V. Adarsh, J. Non-Cryst. Solids **426**, 72 (2015).

(20) P. Khan, A. Joshy, A. Bhattacharya, and K.V. Adarsh, J. Non-Cryst. Solids **449**, 70 (2016).

(21) P. Khan, A. Bhattacharya, A. Joshy, V. Sathe. U. Deshpande, and K. V. Adarsh, Thin Solid Films **621**, 76 (2017).

(22) J. Tauc, R. Grigorovici, A. Vancu, Phys. Status Solidi **15**, 627 (1966).

(23) A. Ganjoo, and R. Golovchak, J. Optoelectron. Adv. M. **10**, 1328 (2008).

(24) A. V. Kolobov and S. R. Elliott, Advs. Phys. **40**, 625 (1991).

(25) S. R. Elliott, J. Non-Cryst. Solids **130**, 85 (1991).

(26) H. Jain, A. Kovalskiy and A. Miller, J. Non-Cryst. Solids **352**, 562 (2006).

(27) M. Popescu, F. Sava, A. Lorinczi, A. Velea, M. Vlcek and H. Jain, Philos. Mag. Lett. **89**, 370 (2009).

(28) F. Sava, M. Popescu, A. Lorinczi and A. Velea, Phys. Status Solidi B **250**, 999 (2013).

(29) G. Kluge, Phys. Status Solidi A **101**, 105 (1987).

(30) A. T. Steel, G. N. Greaves, A. P. Firth and A. E. Owen, J. Non-Cryst. Solids **107**, 155 (1989).

(31) I. T. Penfold and P. S. Salmon, Phys. Rev. Lett. **64**, 2164 (1990).

(32) A. Kovalskiy, A. Ganjoo, S. Khalid and H. Jain, J. Non-Cryst. Solids **356**, 2332 (2010).

(33) V. P. Palumbo, A. Kovalskiy, H. Jain and B. D. Huey, Nanotechnology **24**, 125706 (2013).